# Evaluation of strain and charge-transfer doping in wet-polymeric transferred monolayer MoS$_2$: implications for field effect transistors

C. Abinash Bhuyan[1], Kishore K. Madapu[1], K. Prabhakar[1], Jagnaseni Pradhan[2], Arup Dasgupta[3], S R Polaki[1], Sandip Dhara[1]

[1] Surface and Nanoscience Division, Indira Gandhi Centre for Atomic Research, A CI of Homi Bhabha National Institute, Kalpakkam 603 102, India
[2] Defect and Damage Studies Section, Material Science Group, Indira Gandhi Centre for Atomic Research, Homi Bhabha National Institute, Kalpakkam-603 102, India
[3] Physical Metallurgy Division, Metallurgy and Materials Group, Indira Gandhi Center for Atomic Research, Kalpakkam, Tamil Nadu 603102, India

*Abstract*

Two-dimensional (2D) materials offer exceptional tunability of electronic and optical properties via strain and doping engineering. However, the unintentional introduction of polymeric residues during wet-chemical 2D film transfer processes such as wet-chemical etching and surface-energy-assisted methods remain critical, yet unexplored. This study systematically investigates the impact of such residues on the optical and electrical properties of monolayer MoS$_2$ (1L-MoS$_2$) using Raman and photoluminescence (PL) spectroscopy. We reveal that polymer residues in transferred film from wet-chemical etching induce distinct strain and doping behaviors: PMMA-existed regions exhibit biaxial tensile strain and *p*-type doping, while PMMA-free regions show compressive strain. In contrast, the surface-energy-assisted transfer method introduces compressive strain and *n*-type doping in the transferred film due to residue interactions. Field-effect transistor (FET) measurements corroborate these findings, showing polymer residue-influenced modulation of charge transport. Notably, the surface-energy-assisted technique minimizes transfer-induced defects, highlighting its superiority for fabricating high-performance 2D optoelectronic devices. These results highlight the critical role of transfer methodologies in tailoring optoelectronic properties and provide practical insights for optimizing 2D material integration in advanced technologies.

**Introduction**

Two-dimensional (2D) transition metal dichalcogenides (TMDCs) are recognized for their exotic electronic properties which enable them in various optoelectronic applications. [1, 2] Among 2D TMDCs, the monolayer $MoS_2$ (1L-$MoS_2$) is gaining a lot of attention in the research communities because it possesses a semiconducting nature with direct-bandgap (1.84 eV), the high binding energy of quasiparticles and high absorption coefficient. [1, 2] In the synthesis of large-area monolayer $MoS_2$, chemical vapor deposition (CVD) proves better among 2D film synthesis methods.[3] However, the high growth temperature and harsh chemical environment of CVD restrict the direct synthesis of the monolayers on polymeric substrates like polyethylene terephthalate and polyethylene-naphthalate. [4] To compensate for the high-temperature budget, the as-grown film needs to transfer onto any suitable substrate for various optoelectronic applications. Under fundamental research, the fabrication of vertical heterostructures or moire materials using any 2D TMDCs are the most fascinating research of recent times.[5] In addition, integrating the transferred 2D film with the existing silicon technologies promises to provide enhanced silicon technology.[6] In addition, the film transfer process is ineluctable for the fabrication of various transparent and flexible electronic devices.[6, 7]

In general, the 2D film can be transferred by the wet-polymeric transfer method, dry determinist transfer method and metal-assisted film transfer method. [8] Each film transfer method has its advantage and disadvantages. In the application scenario, the transfer of a large-area film is a prime requirement.[9, 10] In that context, the adoption of the wet-polymeric transfer method is most suitable and preferred over all other existing film transfer methods .[9] However, the only daunting issue in wet-polymeric transfer methods is the existence of polymer residues in the transferred film.[11, 12] The existence of polymer residues can manipulate the optical, and electrical properties of the film.[13, 14] Notably, being a foreign material, the existence of unintentional polymer residues can generate strain and doping.[15] Prior research works have investigated the existence of foreign particles on the surface of 1L-$MoS_2$, which has the potential to generate both strain and surface charge transfer doping (SCTD).[16] As the lattice constant of surface dopant is different from 2D material, this potentially generates strain in the film.[16] In addition, the existence of foreign materials with different work functions as a reference to $MoS_2$ can induce SCTD. [17] The consequences of SCTD in $MoS_2$ can be studied by analyzing both optical and electrical results. Among optical analysis, the manipulation of phonon and electronic properties were well-studied by Raman and PL spectroscopies ,respectively.[16, 18-20] Previously, the optical properties of 2D material are manipulated by the superacid on the $MoS_2$ surface to influence the surface charge transfer.[21] By studying the electrical properties, the variations in electrical conductivity can be manifested in field-effect-transistors (FET) characteristics.[17] In this context, the CVD grown 1L-$MoS_2$ mostly exhibits *n*-type conductivities because of the generation of S vacancies in film synthesis step.[22] The SCTD has immense potential to induce both *n*- and *p*-type doping with varying doping degrees.[17] The existence of noble metals on the surface of $MoS_2$ induces *p*-type doping by depleting the electrons from the $MoS_2$ film. In particular, Pt nanoparticles (NPs) act as a *p*-type dopant in contrast yttrium (Y) NPs act as an *n*-type dopant to the MoS$_2$.[23] Moreover, self-assembled monolayers with different functional groups such as electron-withdrawing (-CH$_3$) or donating (-NH$_3$) groups, act as *p*-type or *n*-type dopants to 1L-MoS$_2$ ,respectively. [24, 25] Furthermore, surface dopants such as K and benzyl viologen (BV) were used as *n*-type surface dopants to 1L-MoS$_2$.[26, 27] To date, all the SCTD methods were intentionally carried out to manipulate the conductivities of pristine MoS$_2$. In contrast, the SCTD due to the left-over polymer residues in transferred 1L-MoS$_2$ hasn't been explored so far. It is worth mentioning that being an organic polymer, left-over polymer residues have the potential to induce strain and surface doping in the transferred flake.

In this report, we comprehensively studied the impact of polymer residues in the wet polymeric transferred 1L-MoS$_2$ flakes. The existence of unintentional left-over polymer residues as polymethyl methacrylate (PMMA) from the wet-chemical etching method and polystyrene (PS) polymer from the surface-energy-assisted transfer method acted as a surface dopant to the transferred 1L-MoS$_2$ flakes. The introduced strain and doping due to residues were decoupled by the Raman correlative plot analysis. Raman results were well supported by the photoluminescence (PL) studies. In addition, the spectroscopic results were further validated by the back-gated FET results.

**Result and discussion:**

The large-area 1L-MoS$_2$ flakes were synthesized via chemical vapor deposition (CVD), as confirmed by optical microscopy and field-effect scanning electron microscopy (FESEM). Figure 1a (inset) displays a representative triangular flake with an edge length of ~10 μm, consistent across optical and FESEM imaging, confirming the uniformity of the CVD-grown structure.

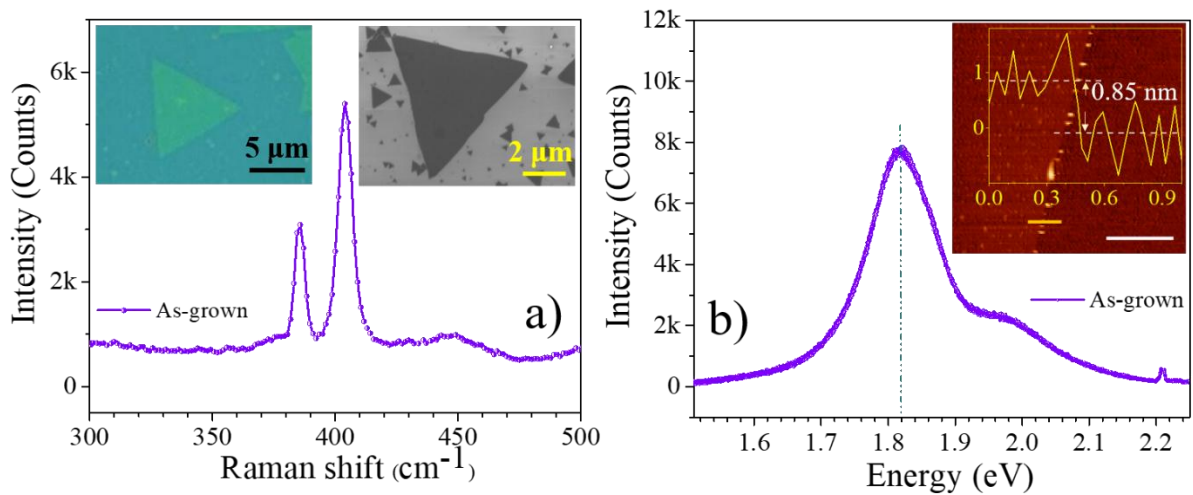

**Figure 1** Confirmation of synthesis of as-grown 1L-MoS$_2$. (a) A representative Raman spectrum of as-grown 1L-MoS$_2$. Insets show the optical and FESEM images of typical triangular flakes. (b) PL spectrum of 1L-MoS$_2$ with intense A-exciton energy at 1.82 eV. Inset shows AFM micrographs with their line profile. Scale bar: 2 μm.

The 2D TMDCs are well characterized by Raman spectroscopy owing to the covalent bonding between the constituent atoms.[28] The 1L-MoS$_2$ shows two distinct Raman modes, assigned as $E^1_{2g}$ phonon mode corresponding to the out-of-phase in-plane vibrations respectively, of Mo and S atoms. $A_{1g}$ phonon modes correspond to out-of-plane vibrations of two S atoms.[29] For statistical Raman analysis, multiple Raman spectra were collected from a triangular flake. Figure 1a shows the typical Raman spectrum of as-grown $MoS_2$ flakes. In the as-grown flakes, the $E^1_{2g}$ and $A_{1g}$ Raman modes were found to be 385.46±0.03 and 404.19±0.02 cm$^{-1}$, respectively. The difference (Δ) in the Raman shift is 18.73±0.22 cm$^{-1}$ which confirms the monolayer thickness of the $MoS_2$ flake. Photoluminescence (PL) spectroscopy further validated the optical quality and electronic structure of the 1L-$MoS_2$. Deconvolution of the PL spectrum (Fig. 1b) into three Gaussian components resolved the A-exciton (1.82 eV), trion (1.80 eV), and B-exciton (2.00 eV) transitions (Fig. 1b and Fig. S1) In general, the characteristics of PL spectra of 1L-$MoS_2$ consist of intense A-exciton (~ 1.84 eV) emission . Therefore, the as-grown 1L-$MoS_2$ flakes shows high optical quality. Atomic force microscopy (AFM) provided direct thickness confirmation, revealing a step height of 0.85 nm at the flake edge (Fig. 1b inset). While slightly larger than the theoretical monolayer thickness (~0.7 nm), this discrepancy is attributed to interfacial interactions with the $SiO_2$/Si substrate, consistent with prior reports [30]. The AFM data, combined with Raman and PL results, conclusively establish the monolayer nature and structural integrity of the synthesized $MoS_2$.

We cut the sample with triangular 1L-$MoS_2$ flakes into two pieces and further, transfer onto two other $SiO_2$/Si substrates using the two most practised wet-polymeric transfer methods: the wet-chemical etching method and the surface-energy-assisted transfer method. Particularly, the PMMA and PS are carrier polymers in the wet-chemical etching method and surface-energy-assisted transfer method , respectively [31, 32] We named the flake transferred by the wet-chemical etching method and surface energy-assisted method as PMMA/1L-$MoS_2$ and PS/1L-$MoS_2$, respectively. As the substrate also induces doping in $MoS_2$ flake,[33] We used another $SiO_2$ substrate (from the same wafer of substrates used for as-grown film) to observe only transfer-induced changes. Notably, we studied the impact of left-over polymer residues in the optical and electrical properties of the transferred flakes.

The detailed film transfer methodologies were discussed in the experimental sections. Figure 2a shows the optical image of transferred PMMA/1L-$MoS_2$ flakes. The optical image illustrates the triangular 1L-$MoS_2$ flakes with few chunks of residues. The observed PMMA residues in transferred 1L-$MoS_2$ were also reported earlier. [12, 14, 34] For a clear observation of the PMMA chunks over the triangular flake, we have carried out the high-magnification FESEM characterisation. The FESEM image confirms the existence of residues-like a chunk on both $SiO_2$ and $MoS_2$ surfaces (Fig. 2b). In addition, the edge and corner of the representative triangular flakes were found to be ruptured in the transferred film. Mostly, such results were common in the PMMA/1L-$MoS_2$.[12] To examine the topography of the residues on transferred film, several AFM measurements were carried out. A typical AFM image delineates the existence of the PMMA residues on the transferred 1L-$MoS_2$ (Fig. 2c). Such results were common in PMMA/1L-$MoS_2$ film.[35] The mean height of the PMMA chunks was calculated by AFM measurement and found to be ~ 80 nm. (Fig. S2) The maximum height of the PMMA chunk is ~120 nm. Likewise, we have also carried out similar measurements in the PS/1L-$MoS_2$ film. (Fig. 2d-f). The optical image shows the damaged- and wrinkle-free triangular flakes. (Fig. 2d) The FESEM image of a typical triangular flake delineates the non-uniformity in surface morphology. (Fig. 2e) A typical high-magnified AFM image represents the surface topography of the PS/1L-MoS$_2$ film. (Fig. S3) Interestingly, the PS/1L-MoS$_2$ flakes also contain nanoparticle-like residues. To the best of our knowledge, the PS residues on transferred film doesn't exist in literature. In literature, there is a tendency to show the low-magnification AFM image for which the existence of PS residues on transferred film is overlooked. [123] However, our high magnified AFM image discernible the PS residues. (Fig 2f and S3). The roughness of the PS/1L-MoS$_2$ film was measured to be ~12 nm. (Fig. S3) The maximum and mean height of PS residues were found to be, ~35 and ~8 nm respectively which covered all the PS/1L-MoS$_2$ flake surfaces.

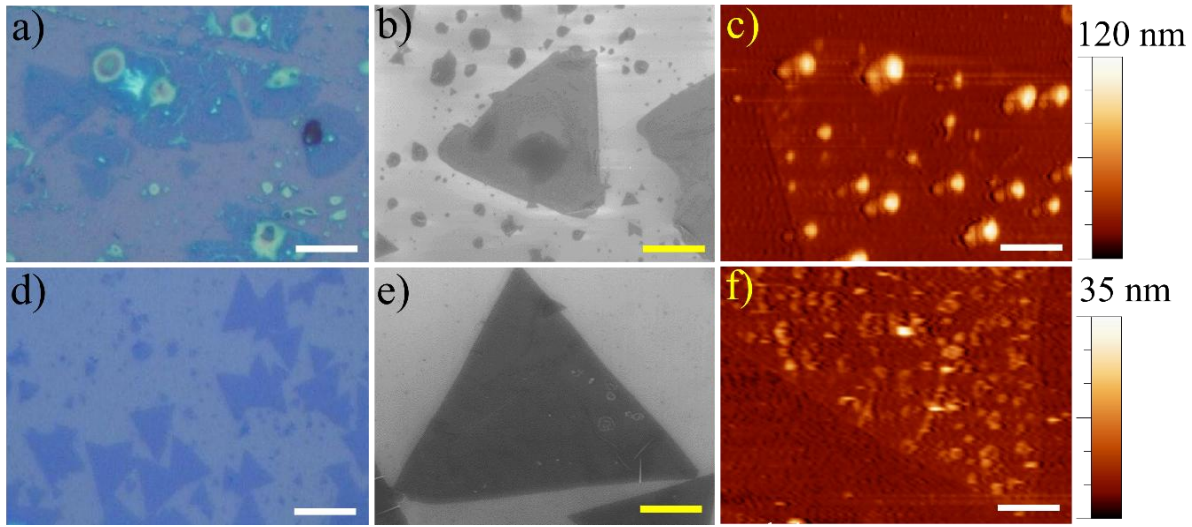

**Figure 2** 1L-MoS2 flakes transferred by (a-c) wet-chemical etching method and (d-f) surface-energy-assisted transfer method. The optical, FESEM and AFM image of 1L-MoS$_2$ flakes by wet-chemical etching and surface-energy-assisted transfer method. Scale bar is 5 μm, 2 μm and 1 μm for optical, FESEM and AFM image respectively.

For nanoscale spatial characterisation, we employed high-resolution transmission electron microscopy (HR-TEM) in wet-polymeric transferred film. A low magnification HR-TEM image of PMMA/1L-MoS$_2$ film depicts the nano-sized PMMA residues present in the film. (Fig. S4a) In Fig. 3a, a high-magnification HR-TEM image of off-PMMA/1L-MoS$_2$ film depicts the strip-like leftovers on the transferred film, which is attributed to the PMMA polymer.[12] Likewise, the PS/1L-MoS$_2$ film shows the nano-sized residues on the film which were recognized as the PS polymer. (Fig. 3b) Hence, it ascertains that the wet-polymeric transferred flakes are unclean at the nanoscale. Moreover, from the fast Fourier transform image, we calculated the *d*-spacing value, 0.27 nm which represents the (100) plane of 2H-MoS$_2$. [12]

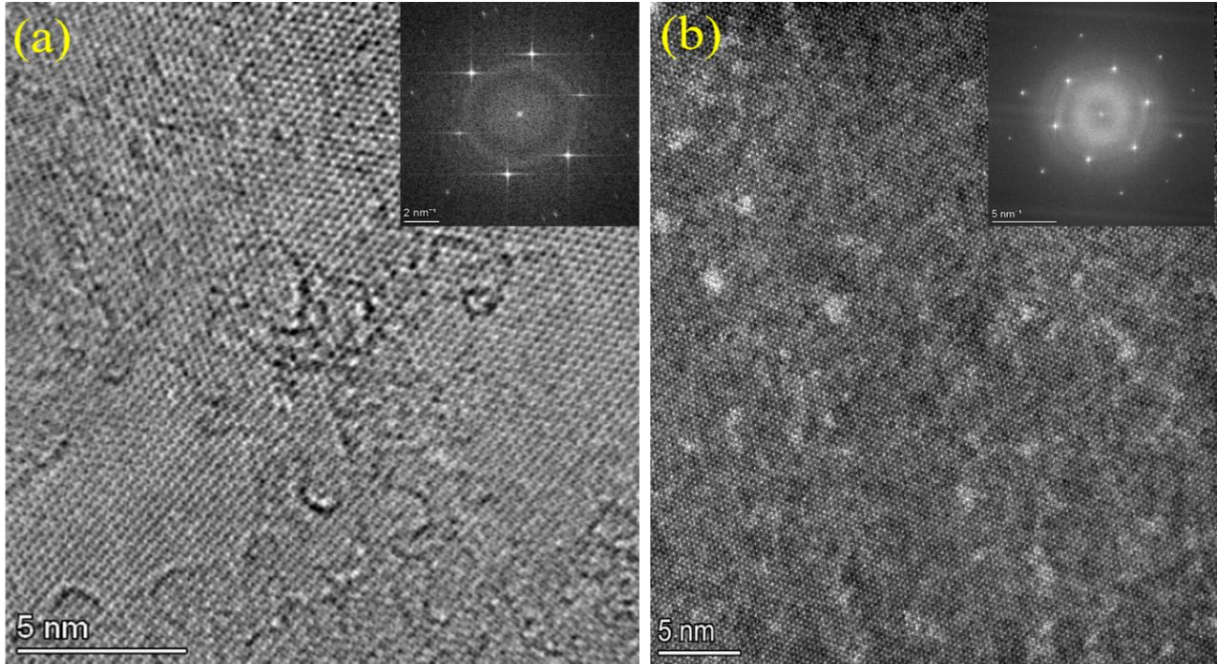

**Figure 3** Atomic resolution HR-TEM image of (a) off-PMMA /1L-MoS$_2$ and (b) PS/1L-MoS$_2$ film showing the surface cleanliness. The corresponding fast Fourier transform image is provided in the inset.

The wet-polymeric transferred 1L-MoS$_2$ samples were assessed by Raman and PL spectroscopies for studying their variation in optical and electronic properties as might be manipulated by existence of the polymer residues on 1L-MoS$_2$ surfaces. In that context, a small micron-sized PMMA residue covering 1L-MoS$_2$ flake was selected and assessed by Raman spectroscopy. The spots with the residues are assigned called on-PMMA/1L-MoS$_2$ and the rest of the flake region is called off-PMMA/1L-MoS$_2$ (insets of Fig. 4a). For relevant statistical calculation purposes, we have also carried out Raman imaging measurements on other PMMA/1L-MoS$_2$ flakes. (Fig. S5) Since the residue size is ~1 μm and the typical flake size is ~10 μm, spectroscopic imaging is advantageous for the fair comparison of the extracted spectra collected from on-PMMA and off-PMMA regions. From Raman imaging dataset, a few representative Raman spectra of on-PMMA and off-PMMA/1L-MoS$_2$ are extracted and presented in Fig 4a and Fig 4b respectively. The Raman intensity of the on-PMMA region is ~2 times higher than the off-PMMA region. The origin of the enhanced Raman intensity may be because of the residue-generated interferences which were explained by the multiple reflection model.[36] The detailed statistical Raman analysis is provided in Tables S1 and S2. In the on-PMMA region, the mean $E^1_{2g}$ and $A_{1g}$ Raman shift are extracted to be 384.93±0.36 and 406.61±0.33 cm$^{-1}$ respectively. Similarly, the mean Raman shift in off-PMMA region was extracted to be 385.60±0.01 and 405.59±0.16 cm$^{-1}$ respectively. In addition, the mean Δ value is found to be 21.68±0.83 and 19.99±0.41 cm$^{-1}$ in on-and off-PMMA/1L-MoS$_2$ region respectively. The mean Δ value is a little higher in on-PMMA than off-PMMA because a red(blue)-shift was observed in $E^1_{2g}$ ($A_{1g}$) mode. The overall increase in Δ value is attributed to a change in strain and doping in the transferred 1L-MoS$_2$.[15] Unlike Raman intensity, Raman shift reveals the residue-induced strain and doping due to SCTD.[37] In literature, red-shift in $E^1_{2g}$ mode is attributed to tensile strain and blue-shift in $A_{1g}$ mode is due to *p*-type doping of 1L-MoS$_2$.[15] The observed Raman shift in both the modes used for the strain and doping calculations is explained in little later. The observed Raman result can be solidified in Raman imaging measurements. The colour map corresponds to Raman intensity which showed a higher value at the residue region (on-PMMA) (Fig. 4 c) as compared to rest of the flake area region (off-PMMA) (Fig.4e). Moreover, the red-shift in $E^1_{2g}$ mode (Fig. 4d) and blue-shift in $A_{1g}$ mode is observed (Fig. 4f).

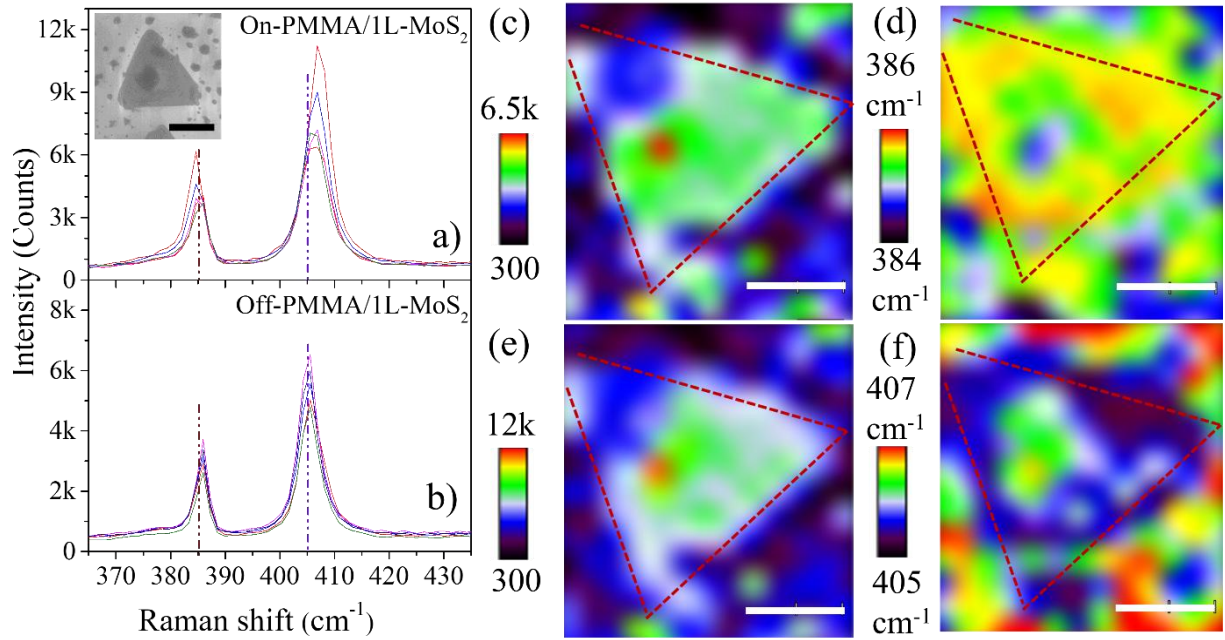

**Figure 4** Raman imaging analysis of wet-polymeric transferred 1L-MoS$_2$ film. Representative Raman spectra of (a) on-(b) off-PMMA chunk on 1L-MoS2. Inset shows the FESEM image of selected triangular flake for Raman imaging. Scale bar: 5 μm. A colour map was illustrated using Raman imaging which are representing the (c) intensity (d) position of E$^1_{2g}$ mode. (e) A colour map was drawn using Raman imaging which represents the (c) intensity (d) position of A$_{1g}$ mode. The red dashed triangular outlines on the colour map are guiding the eyes. Scale bar: 5 μm.

Similarly, Raman imaging of the PS/1L-MoS$_2$ film was also carried out. (Fig 5) Few representative spectra were extracted arbitrarily from all the datasets and plotted in Fig. 5a. Both Raman modes show the invariation in peak position and intensity (Fig. 5a). In the PS/1L-MoS$_2$ flake, the mean $E^1_{2g}$ and $A_{1g}$ Raman shifts are found to be 385.64±0.05 (Fig. 5c) and 403.32±0.08 cm$^{-1}$ (Fig. 5e), respectively. The statistical Raman analysis is provided in Table-S1 and S2. In addition, the Δ value found to be 17.68±0.36 cm$^{-1}$ in PS/1L-MoS$_2$. The negligible blue shift in $E^1_{2g}$ mode may be due to the generated compressive strain due to the existence of PS polymer residues and primarily responsible for the observed reduction in Δ value.[15] The observed Raman shift in both modes calculate the strain and doping which is discussed in little later. For relevant statistical calculation purposes, we have also carried out Raman imaging measurements on other PS/1L-MoS$_2$ flakes. (Fig. S6) Raman intensity colour maps for both $E^1_{2g}$ and $A_{1g}$ are provided in Fig.5b and 5e respectively. Moreover, Raman peak position colour maps of $E^1_{2g}$ and $A_{1g}$ modes are provided in Fig.5c and Fig.5e respectively. The unchanged Raman intensity and peak position delineates the spatial uniformity in the optical properties of PS/1L-MoS$_2$.

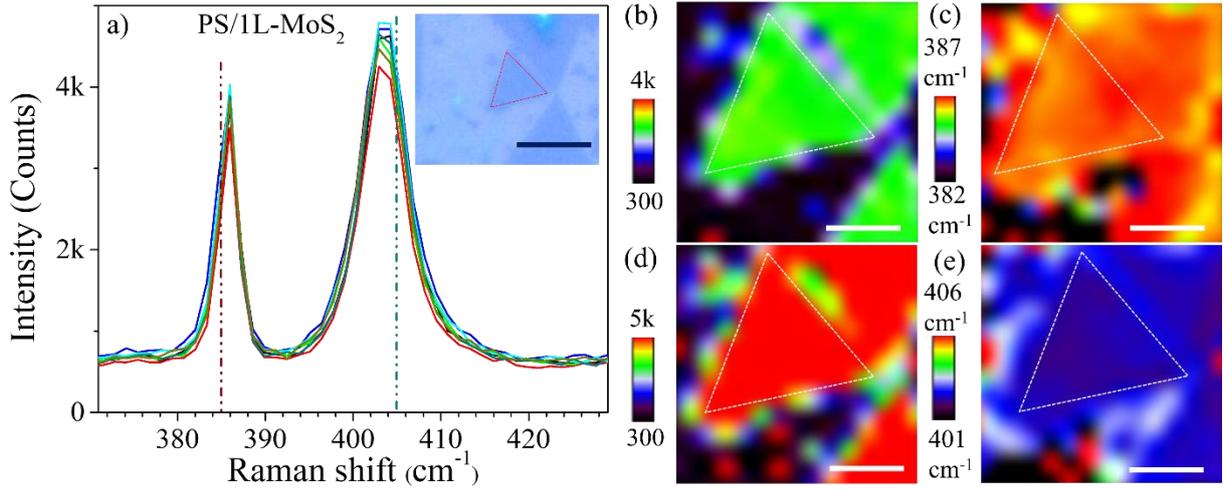

**Figure 5** Raman results of PS transferred 1L-MoS$_2$ film. Representative Raman spectra of (a) PS transferred 1L-MoS$_2$. Inset showing the optical image of a typical triangular flake which Raman imaging was carried out. Scale bar: 10 μm. A colour map was drawn using Raman imaging which represents the (b) intensity (c) position of $E^1_{2g}$ mode. A colour map was drawn using Raman imaging which represents the (d) intensity (e) position of $A_{1g}$ mode. The white dashed triangular outlines on the colour map are guiding the eyes. Scale bar: 5 μm.

The evaluation of the generated strain (ε) and carrier concentration (n) in the transferred 1L-MoS$_2$ is carried out quantitatively using Raman correlative plot.[15] In brief, the correlative plot considers both $E^1_{2g}$ and $A_{1g}$ modes to measure ε and n.[38] In the correlative plot, both Raman modes are represented by red dots for on-PMMA/1L-MoS$_2$, violet dots for off-PMMA/1L-MoS$_2$, and, blue dots for PS/1L-MoS$_2$ films. (Fig. 6) The green-coloured dashed line represents the purely strain axis, the brown-coloured dashed line represents the doping axis and their intersecting point (origin) represents the strain-free and doping-free value for 1L-MoS$_2$ under 532 nm laser excitation. Moreover, each dotted parallel lines correspond to certain value of strain and doping to their corresponding lines.[15] In our analysis, strain-free and doping-free values correspond to the free-standing 1L-MoS$_2$ film which nullifies any substrate effect.[15] The intersecting point is called the origin of the ε-n plot. From the origin, two arrow marks represent the increase in compressive strain and p-type doping in the 1L-MoS$_2$.[15] The dashed brown (green) lines parallel to the corresponding doping-free (strain-free) line serve to quantify the increase with a unit of ±×10$^{13}$ cm$^{-2}$ variation in carrier concentration and ±0.1% of mechanical strain. Each Raman shift value has been incorporated in the following equation (1 and 2) to calculate the ε and n, using the following relationships [15] and presented in the correlative plot.

$$\varepsilon = \frac{k(1)\Delta\omega(2) - k(2)\Delta\omega(1)}{2\gamma(1)\omega(E)k(2) - 2\gamma(2)\omega(A)k(1)} \quad (1)$$

$$n = \frac{\gamma(1)\omega(E)\Delta w(2) - \gamma(2)\omega(A)\Delta w(1)}{\gamma(1)\omega(E)k(2) - \gamma(2)\omega(A)k(1)}, \quad (2)$$

where $k(1)$ and $k(2)$ are proportionality constants of the relationship between peak frequency and carrier concentration (n). The $k(1)$ and $k(2)$ values are -0.33× 10$^{-13}$ and -2.22× 10$^{-13}$ for $E^1_{2g}$ and $A_{1g}$ Raman modes, respectively. ω(E) and ω(A) represent the strain-free and doping-free values of $E^1_{2g}$ (385 cm$^{-1}$) and $A_{1g}$ (405 cm$^{-1}$) Raman modes for 1L-MoS$_2$, respectively. Δω(1) and Δω(2) are Raman shifts in $E^1_{2g}$ and $A_{1g}$ modes with respect to the strain-free and doping-free values. γ(1) and γ(2) are room temperature Gruneisen parameters of $E^1_{2g}$ (0.86) and $A_{1g}$ (0.15) Raman mode, respectively [15, 37].

Figure 6 represents ε-n plot which contains the extracted Raman shift values. In on-PMMA/1L-MoS$_2$, the film undergoes tension and the % of tensile strain varies from ~0.01 to ~0.17% with a mean value of ~0.07%. Large polymer residues undergo significant volumetric shrinkage as solvents evaporate or during thermal curing. This shrinkage exerts a pulling force on the underlying MoS$_2$ lattice, stretching it outward and inducing tensile strain.[39] Like strain, we also evaluated the SCTD in the on-PMMA region. The calculated n value confirmed the p-type doping of 1L-MoS$_2$ and the hole concentration was varied from ~ $0.52\times10^{13}$ to ~$1.09\times10^{13}$ cm$^{-2}$ with the mean value of ~$0.74\times10^{-13}$ cm$^{-2}$. The induced p-type doping is due to the transfer of electrons from 1L-MoS$_2$ to PMMA residue. In contrast, the off-PMMA/1L-MoS$_2$ undergoes compression, and the mean compressive strain is ~0.11%. Like the on-PMMA region, the off-PMMA region also undergoes p-type doping with a hole concentration, of ~$0.19\times10^{13}$ cm$^{-2}$. However, the hole doping concentration is lower in off-PMMA/1L-MoS$_2$ as compared to on-PMMA/1L-MoS$_2$. The observed compressive strain and negligible p-type doping in off-PMMA/1L-MoS$_2$ were attributed to the existence of the irregularly scattered nano-sized PMMA residues which was shown in the HR-TEM image. In general, the CVD grown 1L-MoS$_2$ films are compressively strained due to high growth temperature.[40] In our as-grown 1L-MoS$_2$, the film undergoes compressive strain and n-type doping concentration with ~ 0.12±0.01 % and ~ $0.46±0.02\times10^{-13}$ cm$^{-2}$ respectively. Therefore, the result is quite obvious for CVD-grown samples and also reported by earlier studies.[41] To sum up, the existence of left-over PMMA residues induces tensile strain and p-doping to 1L-MoS$_2$.

Similarly, the ε and n of PS/1L-MoS$_2$ were evaluated using Raman correlative plot. The PS residues generate % of compressive strain, ~ 0.17±0.01 % and increase in the electron carrier concentration, ~$0.89±0.04\times10^{13}$ cm$^{-2}$ in 1L-MoS$_2$ film (Table-1). For comparison, the as-grown 1L-MoS$_2$ film shows compressive strain, 0.12±0.01% and electron concentration, $0.46±0.02\times10^{13}$ cm$^{-2}$ (Table-1). Therefore, the PS/1L-MoS$_2$ undergoes higher compressive strain because of the existing nano-sized polymer residues. Since the size of the residue is small (mean roughness value is ~12 nm in AFM result), the PMMA may be bond to specific sites on MoS$_2$, such as sulfur vacancies or edges. These localized interactions can distort the lattice inward, creating compressive strain. The nano-sized PS residues are also observed in HR-TEM results. (Fig.3b) In addition, the n-type doping density also increased in PS/1L-MoS$_2$ as compared to as-grown film. The observed additional n-type doping is attributed to the transfer of the electrons from PS residues to 1L-MoS$_2$ film, raising fermi level of MoS$_2$ toward its conduction band [42]. Since polystyrene has a lower work function (Φ_polymer < Φ_MoS$_2$),[42] leading to electron transfer upon contact. Therefore, PS/1L-MoS$_2$ behaves as n-type doped in contact with PS polymer.

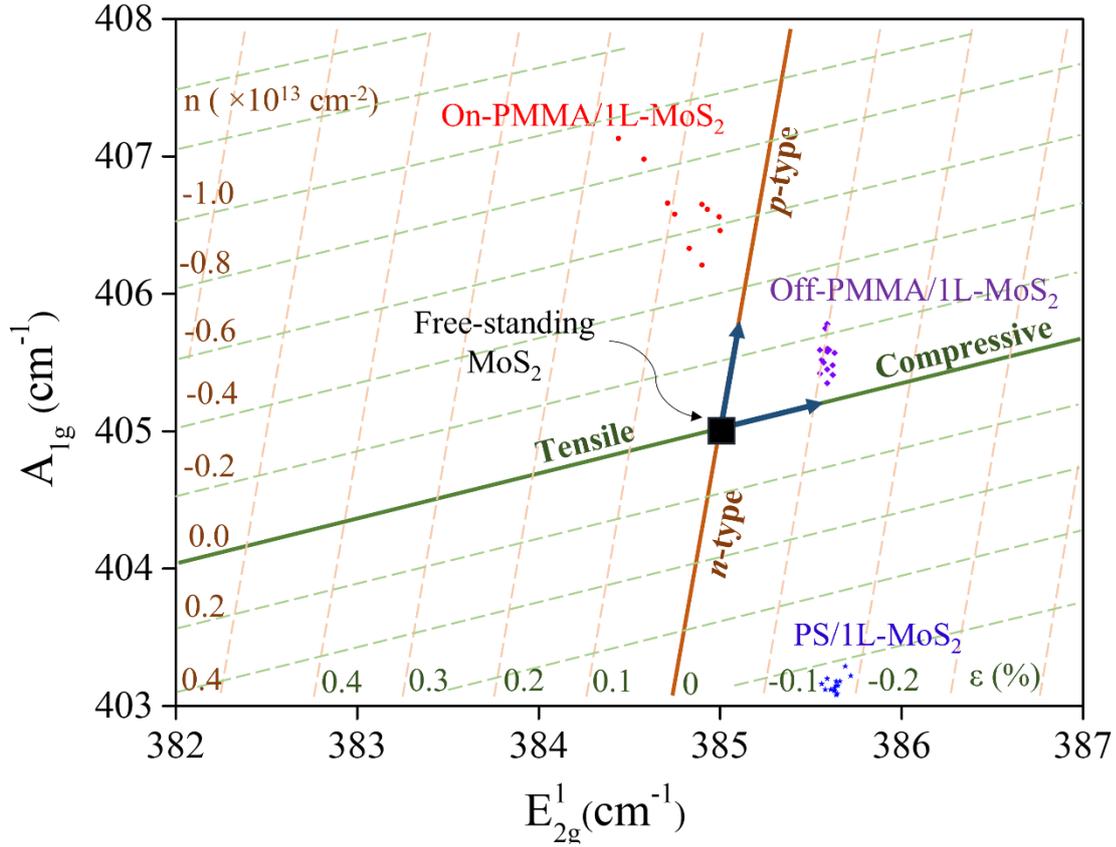

**Figure 6** Correlation plot of $E^1_{2g}$ and $A_{1g}$ Raman modes corresponds to strain and doping distributions in wet-polymeric transferred 1L-MoS$_2$.

The statistical Raman analysis have been tabulated with Raman peak positions, generated strain, and induced carrier concentrations. (Table-1 and S3)

**Table 1:** Statistical analysis of strain and doping concentrations calculated from the observed Raman shifts.

| Sample | Mean E-peak | Mean A-peak | Δ | ε (%) | Strain type | $n$ (×10$^{13}$) cm$^{-2}$ | Doping type |
|---|---|---|---|---|---|---|---|
| As-grown | 385.46±0.03 | 404.19±0.02 | 18.73±0.22 | -0.12±0.01 | Compressive | 0.46±0.02 | *n*-type |
| On-PMMA | 384.93±0.36 | 406.61±0.33 | 21.68±0.83 | 0.07±0.02 | Tensile | -0.78±0.21 | *p*-type |
| Off-PMMA | 385.60±0.01 | 405.59±0.16 | 19.99±0.41 | -0.10±0.01 | Compressive | -0.19±0.08 | *p*-type |
| PS | 385.64±0.05 | 403.32±0.08 | 17.68±0.36 | -0.17±0.01 | Compressive | 0.89±0.04 | *n*-type |

To further elucidate the impact of polymer residues on the electronic properties of the transferred film, we carried out the PL measurements (Fig. 7). [8] Notably, the on-PMMA/1L-MoS$_2$ showed lower overall intensity as compared to off-PMMA/1L-MoS$_2$ region. The lowering in PL intensity at the residue region can be attributed to the biaxial strain [34], or dielectric screening of the 1L-MoS$_2$ film[36]. From the PL imaging dataset, we segregated to plot the on-PMMA 1L-MoS$_2$ (Fig. 7a) and off-PMMA 1L-MoS$_2$ (Fig. 7b). The representative spectrum fitted with the Gaussian function to extract the emission energy of trions, A-exciton and B-exciton. [43] By fitting the PL spectra of the on-PMMA region, the emission energy of trions (~1.80 eV), A-exciton (~1.84 eV) and B-exciton (~2.0 eV) were extracted. Moreover, the PL spectra of the off-PMMA region were fitted to extract the emission energy of trions (~1.80 eV), A-exciton (~1.82 eV) and B-exciton (~2.0 eV). An observed negligible blue shift (~20 meV) in A-exciton emission energy at on-PMMA/1L-MoS$_2$ was attributed to the *p*-type doping of the film. Generally, the insulator likely introduces *p*-type doping via charge transfer or fixed negative charges in the insulator. This shifts the Fermi level toward the valence band, filling low-energy states which resulting increases the effective bandgap as Burstein-Moss effect [44]. Moreover, it may be contributed by the increased dielectric screening because the dielectric constant ($\kappa$) of the polymer residues ($\kappa_{PMMA}$ ~ 3 and $\kappa_{PS}$ ~ 2.6) is higher than air. [45] In a nutshell, the observed PL result is well supported by Raman results.

To show the spatial variation in PL emission, we solidify our Raman analysis results with PL imaging. The overall PL intensity and peak energy of A-exciton (1.82 eV) were provided as colour maps (Fig. 7 b-e). The overall PL intensity of PMMA/1L-MoS$_2$ (Fig. 7b) was higher than PS/1L-MoS$_2$ (Fig. 7d) film. Notably, PS/1L-MoS$_2$ shows emission peak at 1.82 eV as as-grown 1L-MoS$_2$ flakes. In addition, A-exciton peak energy (1.82 eV) was found blue-shifted in PMMA/1L-MoS$_2$ as compared to PS/1L-MoS$_2$ (Fig. 7d and 7e). The linewidth of A-exciton in PS/1L-MoS$_2$ (~180 meV) and PMMA/1L-MoS$_2$ (~130 meV) was extracted by Gaussian fitting of PL spectra. The increase in the PL linewidth can also confirm the *n*-type doping of the PS/1L-MoS$_2$ film.[8] As a result, we can safely conclude that the PS/1L-MoS$_2$ film becomes *n*-doped whereas PMMA/1L-MoS$_2$ becomes *p*-type doped when respective residues have existed over the film.

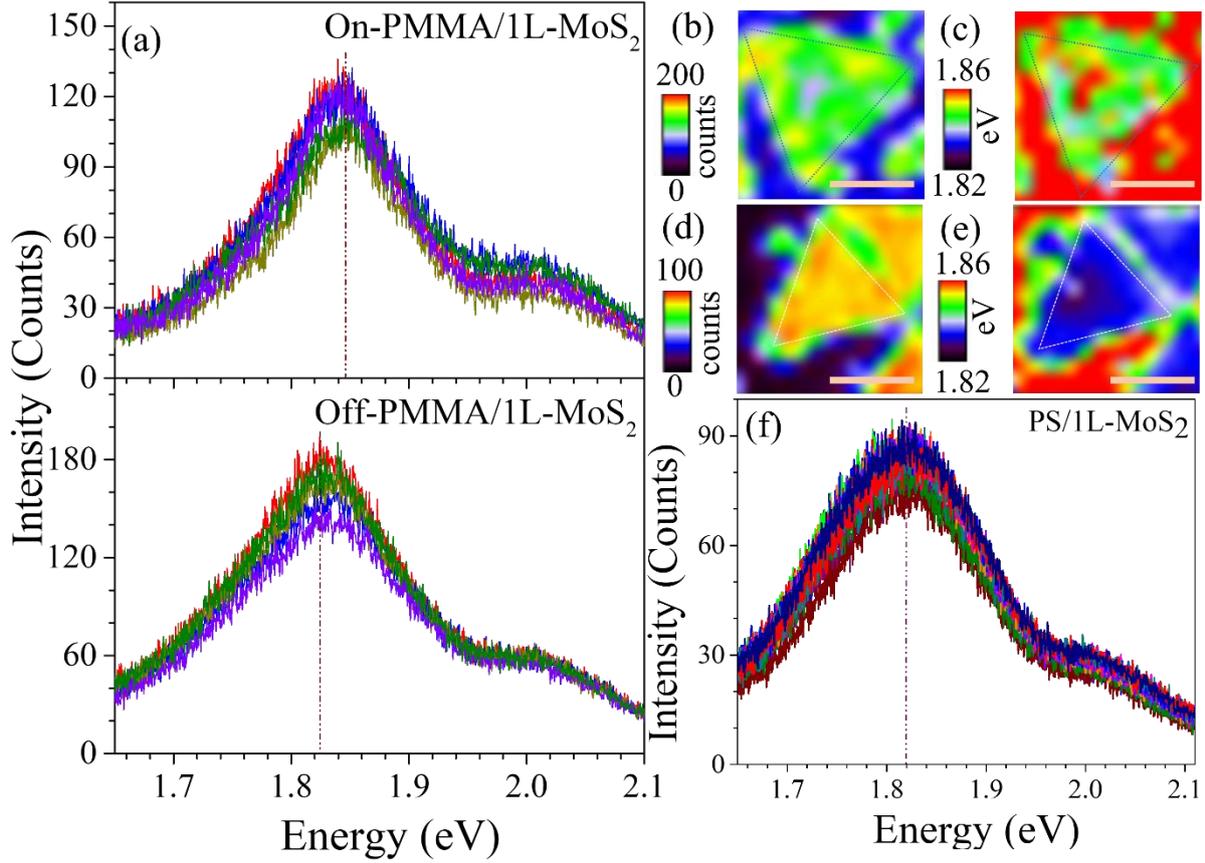

**Figure 7** PL imaging of the as-selected triangular flake. For PMMA/1L-MoS$_2$ film: (a) the extracted PL spectra from the PL imaging dataset, colour map of (b) overall PL intensity (c) A-exciton peak energy. For PS/1L-MoS$_2$ film: colour map of (d) overall PL intensity (e) A-exciton peak energy, (f) the extracted PL spectra from the PL imaging dataset. Scale bar: 5 μm.

To evaluate the presence of C concentration in transferred 1L-MoS$_2$ film, we employed Rutherford back-scattering spectroscopy (RBS) technique, at the carbon resonance energy, 4.2 MeV. The measurement details were provided in the experimental section. Fig 8 represents RBS spectra of polymeric transferred 1L-MoS$_2$ and bare SiO$_2$/Si substrate, as a reference. The inset confirmed the MoS$_2$ phase formation.[46] As the size of the residue is far smaller than the spot size of RBS, ~ 300 μm and also, scattered on film non-uniformly therefore, we are unable to measure the exact C thickness on the transferred film. However, the C content in PMMA/1L-MoS$_2$ is comparatively higher than PS/1L-MoS$_2$ film. From spectroscopic studies, it concludes that the strain and doping density is also higher in PMMA/1L-MoS$_2$ over PS/1L-MoS$_2$. Therefore, we can safely conclude that the amount of C concentration plays an important role in manipulating the strain and doping density in transferred 1L-MoS$_2$ film.

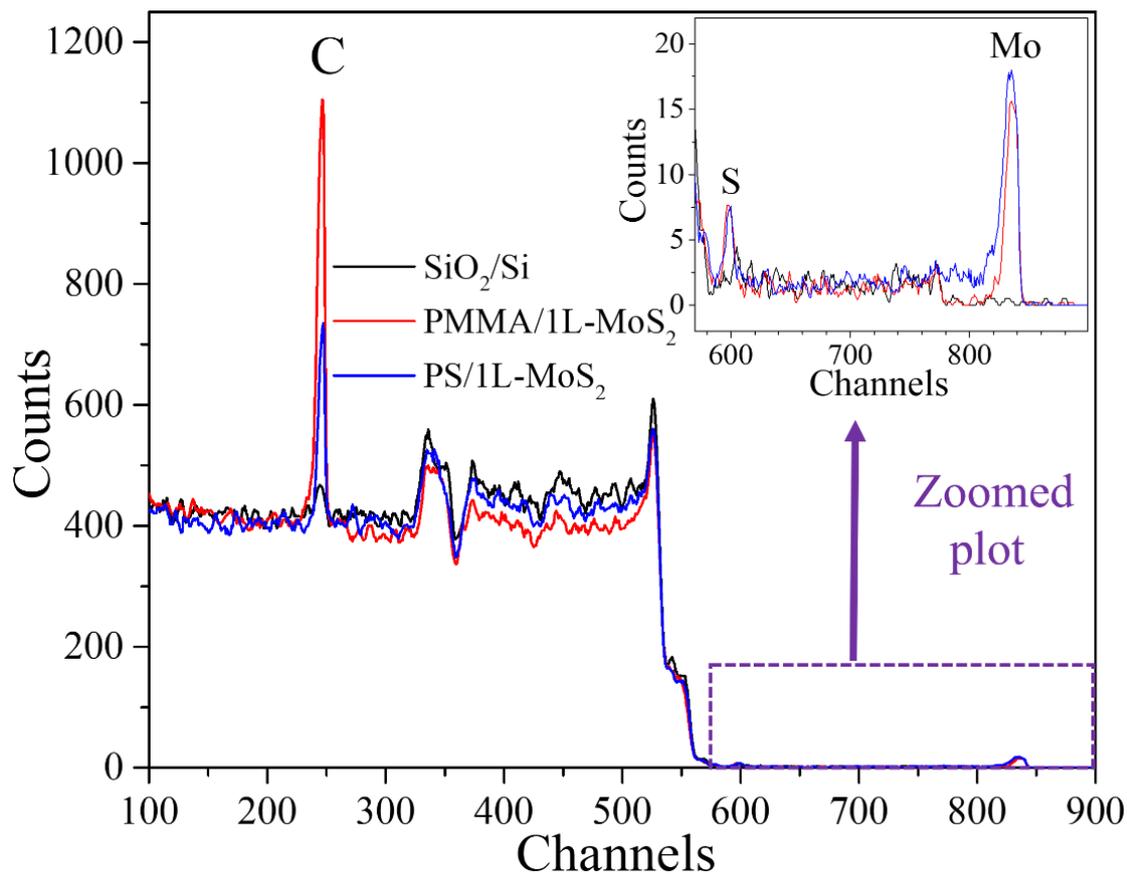

**Figure 8** RBS spectra of the polymeric transferred 1L-MoS$_2$. The bare SiO$_2$/Si substrate is used as a reference.

The influence of polymer residues on the electrical properties of transferred 1L-MoS$_2$ was investigated through the fabrication of back-gated FETs, where the 1L-MoS$_2$ served as the channel material.[17] Detailed experimental protocols for device fabrication and characterization are provided in the Methods section. Schematics of the FET configurations and representative FESEM images of the fabricated devices are shown in Fig. S7. Output and transfer characteristics were systematically analyzed for multiple FETs fabricated using polymeric-transferred 1L-MoS$_2$ films (Fig. S8 and S9). Figure 9a and 9b present the transfer characteristics of typical FETs based on PMMA/1L-MoS$_2$ and PS/1L-MoS$_2$ films, respectively. Both linear and logarithmic scales are included in the graphs to facilitate the extraction of critical device parameters, threshold voltage ($V_t$), sub-threshold swing (SS), ON-OFF current ratio and electron mobility ($\mu$).[2] Additionally, output characteristics (Fig. 9c and 9d) were analyzed to determine the saturation drain current of the FETs. These measurements collectively highlight the role of polymer residues in modulating the electrical performance of 1L-MoS$_2$-based devices.

The transfer characteristics of the FETs, defined $I_{ds}$ versus $V_{gs}$ at constant $V_{ds}$, 1V, are shown in Fig. 9a,b. Both linear and logarithmic scales were used to extract key device parameters. The threshold voltage ($V_t$), determined via the Y-function method [2], was found to be +4.2 V for PMMA/1L-MoS$_2$ and −8.5 V for PS/1L-MoS$_2$. These values indicate p-type doping (positive $V_t$) for PMMA and n-type doping (negative $V_t$) for PS, consistent with prior reports of PMMA-induced p-doping[12] and PS-induced n-doping [42]. This doping behavior is attributed to nano-sized polymer residues introduced during the wet-polymeric transfer process. The sub-threshold swing (SS), which quantifies trap states by measuring the gate voltage required for an order-of-magnitude change in $I_{ds}$ [47], yielded mean values of 1500 mV/decade (PMMA/1L-MoS$_2$) and 1250 mV/decade (PS/1L-MoS$_2$). Both values deviate significantly from the ideal SS of 60 mV/decade, highlighting the role of polymer residues in introducing trap states. The ON/OFF current ratios were $10^5$ (PMMA/1L-MoS$_2$) and $10^6$ (PS/1L-MoS$_2$), with the latter's higher ratio attributed to its enhanced on-state conductivity. Carrier mobility (μ) was calculated using the *eq 3*,

$$\mu = \frac{L}{W \cdot C_{ox} \cdot V_{ds}} \left(\frac{\Delta I_{ds}}{\Delta V_{gs}}\right) \qquad (3)$$

where $C_{ox}$ is the gate insulator capacitance per unit area (11.51 × 10$^{-9}$ F·cm$^{-2}$ for SiO$_2$ of 300 nm), L (10 μA) and W (100 μA) are the effective dimensions of the channel. At $V_{ds}$ of 1V, the slope $\Delta I_{ds}/\Delta V_{gs}$ is calculated from the transfer characteristics curve in a linear regime.[2] The mean carrier mobility is calculated to be 9.5 and 10.3 cm$^2$ V$^{-1}$ s$^{-1}$, for PMMA and PS/1L-MoS$_2$ respectively. A similar range of carrier mobility was also reported in the case of top contact as-grown 1L-MoS$_2$ FETs and typically reported electron mobilities were 1.3 × 10$^{-2}$, 0.005-0.1, and 0.02 cm$^2$ V$^{-1}$ s$^{-1}$.[12, 48] In literature, higher carrier mobility is also reported with devices of similar physical parameters with top-gated configurations. Indeed, the lower FET mobility with CVD-grown film is due to the existence of contaminants, defects, and grain boundaries.[48]

Output FET characteristics define the drain current ($I_{ds}$) versus drain voltage ($V_{ds}$), which are plotted with the variation of gate voltage ($V_{gs}$) from 0 V to 20 V with an interval of 5 V. In both the cases, the transferred film clearly show the FET characteristics i.e., the linear rise in drain current at low voltage (up to ~ 4 V) and further, saturates. Importantly, the perfect current saturation was not observed in both the films which may be attributed to the existence of the respective polymer residues in the transferred film.[12] The saturation in drain current is measured to be ~15 and ~ 25 μA for PMMA (fig. 9c) and PS/1L-MoS$_2$ (Fig. 9d) film respectively. The higher ON state current in PS/1L-MoS$_2$ is attributed to the better channel conductivity as compared to PMMA/1L-MoS$_2$ film. The overall drain current is low in the wet-polymeric transferred film of similar specifications by other studies because of the bottom contact FET configuration.[12] Importantly, the bottom contact FET configuration is advantageous because it is absolutely free from any lithographic polymer residues which generally introduces during the electrode fabrication process.[49-51] As the demand for complementary *p*-type doping and ambipolar doping are useful in the fabrication of logic circuits and *p-n* junction devices [17] Therefore, the selection of the suitable transfer method based on the requirements can be imperative.

**Table 2:** Summarising the key FET device parameters using transferred 1L-MoS$_2$ film as channel material.

| Device parameters | PMMA/1L-MoS$_2$ sample | PS/1L-MoS$_2$ sample |
|---|---|---|
| Threshold voltage ($V_t$) | 4.2± 0.3 V | -8.5 ± 0.2 V |
| Sub-threshold swing (SS) | 1500 ± 120 mV/decade | 1250 ± 100 mV/decade |

| ON/OFF current ratio | $10^5$ | $10^6$ |
| --- | --- | --- |
| Electron mobility (μ) | $9.5 \pm 0.6$ cm$^2$ V$^{-1}$ s$^{-1}$ | $10.3 \pm 0.3$ cm$^2$ V$^{-1}$ s$^{-1}$ |
| Saturation drain current | 15 μA | 25 μA |

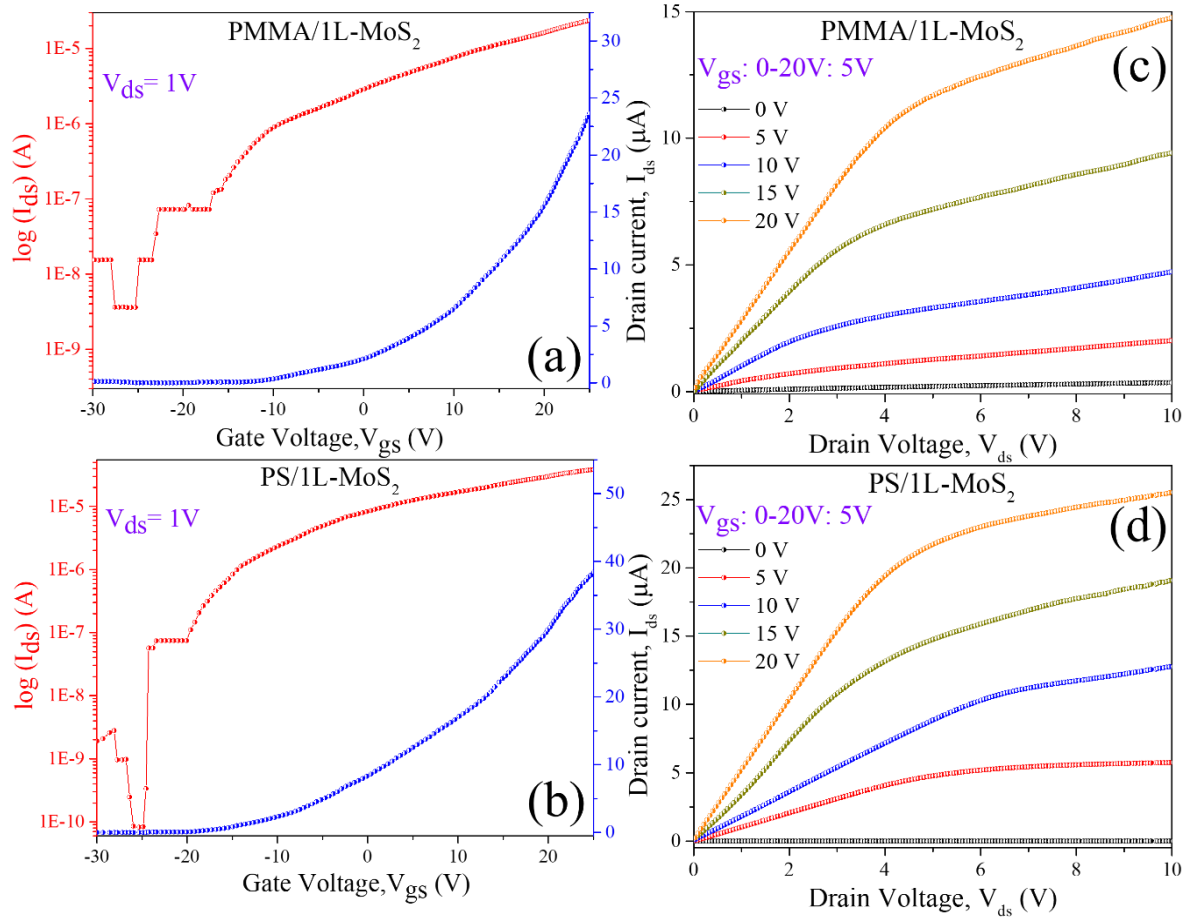

**Figure 9** Transfer and output characteristics of FET by using wet-polymeric transferred 1L-MoS$_2$ film. Transfer characteristics of FET using (a) PMMA/1L-MoS$_2$ film and (b) PS/1L-MoS$_2$ film in both linear and corresponding logarithm scales. Output characteristics of FET using (c) PMMA/1L-MoS$_2$ and (d) PS/1L-MoS$_2$ film.

## Conclusion

Our comprehensive study sheds light on the manipulation of the optical and electrical properties of the wet-polymeric transferred 1L-MoS$_2$ film due to the existence of the left-over polymeric residues. The existence of the residues was confirmed by FESEM, AFM and HR-TEM results. We evaluated the biaxial strain and coupling-induced surface charge-transfer doping in transferred 1L-MoS$_2$ film by using Raman and PL spectroscopies. From the correlative Raman plot analysis, we disentangled the strain and doping contributions. We also noted that the existence of residues of each transferred technique exhibits different types and degrees of strain and doping. The transferred film by wet-chemical etching method undergoes *p*-type doping with on-PMMA (off-PMMA) region and undergoes tensile (compressive) strain. In contrast, the film transfer by surface-energy-assisted method shows compressive strain and *n*-type doping. Our spectroscopic results are well supported by the FET characteristics. FET measurements validate these trends, with PMMA-transferred films showing *p*-type behavior (threshold voltage +4.2 V) and PS-transferred films demonstrating *n*-type characteristics (threshold voltage −8.5 V). Critically, the surface-energy-assisted method minimizes structural defects, yielding devices with superior performance metrics, including enhanced ON/OFF current ratios ($10^6$) and higher electron mobility (10.3 cm² V$^{-1}$ s$^{-1}$). Notably, the surface-energy-assisted transfer process is better than the wet-chemical etching method to fabricate efficient 2D optoelectronic devices.

## Experimental Methods

**Synthesis of 1L-MoS$_2$**

1L-MoS$_2$ samples were synthesized using the atmospheric pressure chemical vapour deposition technique. Molybdenum trioxide (MoO$_3$, 99.97% Sigma Aldrich) and S ( ≥99.5% pure; Sigma Aldrich) powders were used as precursor materials. Details of the growth process may find elsewhere.[41] The source materials were loaded in a three-zone furnace with having one-inch quartz tube as a reaction chamber. The ultra-high pure (UHP) Ar gas was used as the carrier gas and the flow rate of the carrier gas was controlled by a mass flow controller. The polished side of SiO$_2$(300nm)/Si is kept face down over the MoO$_3$ powder in the alumina boat. The optimised amount of S (40 mg) and MoO$_3$ (15 mg) was kept in the first and third zone, respectively. Before starting the growth, the chamber was evacuated to a base pressure of 1 ×10$^{-3}$ mbar and purged UHP Ar with a flow rate of 150 sccm for 15 min. The set temperatures of the first, second and third zones were 160, 200 and 700 °C, respectively. However, the ramp rate of the three zones was adjusted such that the reaching time was the same for all zones. The ramping time of all the zones was 40 mins. The optimised growth time and flow rates were 20 min and 50 sccm, respectively. The furnace was allowed to cool to room temperature naturally after the growth.

**Wet-polymeric transferred 1L-MoS$_2$ onto SiO$_2$(300 nm)/Si substrate**

We have adopted two most extensively practised methods, to evaluate strain and charge-transfer doping. Two different methods: wet-chemical etching [10] and surface-energy-assisted transfer method [32]. The steps of each approach are explained in the following section.

**Wet-chemical etching transfer method:** The conventional wet transfer technique is previously reported.[10] However, we briefly outline the steps in the transfer process. 100 mg of PMMA crystals (MW-80,000 g/mol, Sigma Aldrich) were mixed with 1 ml of anisole (Merck) to prepare the PMMA solution. Subsequently, the prepared PMMA solution was spin-coated over the 1L-MoS$_2$/SiO$_2$(300 nm)/Si substrate with 4000 RPM for 30 sec and subsequently, the substrate was baked for 30 min at 60°C. In the next step, the polymer capped PMMA/1L-MoS$_2$/SiO$_2$(300 nm)/Si substrate was dipped in KOH (1M) solution for 30 min. The bubbles formed on the SiO$_2$ substrate helped to lift off the PMMA-capped monolayer. Then, PMMA capped 1L-MoS$_2$ was fished out by another cleaned SiO$_2$/Si substrate, and which was followed by baking at 90 °C for 30 min for good adhesion. After the baking process, the removal of PMMA films was carried out by rinsing with the acetone several times. Then, the transferred film was annealed at 100 °C under UHP Ar atmosphere for 2 hr.

**Surface-energy-assisted transfer method:** The conventional technique is previously reported.[32] In this process, 180 mg of polystyrene crystals (MW-280, 000 g/mol, Sigma Aldrich) were mixed in 2 ml toluene (Merck) to form the uniform PS solution. Then, the PS solution was spin-coated on 1L-$MoS_2$/$SiO_2$(300 nm)/Si substrate at 3500 RPM for 60 s. Subsequently, the PS-capped monolayer was baked at 90°C for 15 min for good adhesion. After the baking process, PS capped monolayer was dipped in the distilled water and the water was allowed to enter between PS capped monolayer and $SiO_2$/Si substrate by poking a sharp needle near the edge region. Because of the hydrophobic nature of PS, water penetrated between the $SiO_2$ and PS which subsequently helped in the lift-off of the 1L-$MoS_2$ capped with PS. The floating 1L-$MoS_2$ capped with PS film was fished out onto a newly cleaned $SiO_2$ substrate. In the next step, the $SiO_2$ substrate with PS capped 1L-$MoS_2$ was baked at 100°C for 5 min. Subsequently, the PS was removed by dissolving it in toluene. Then, the transferred film was annealed at 100°C under UHP Ar atmosphere for 2 hr.

**Optical spectroscopic characterization:**

Raman and PL analysis was carried out using the micro-Raman spectrometer (inVia, Renishaw, UK). The wavelength of excitation was 532 nm and scattered light was collected in backscattering geometry. All Raman and PL spectra were collected by a thermoelectrically cooled charge-coupled device detector after dispersing through 2400 gr/mm and 1800 gr/mm gratings, respectively. Raman imaging and PL imaging were carried out with 4 sec acquisition time with laser power, ~ 10 μW. The PL imaging was carried out by using grating, 600 gr/mm.

**AFM and FESEM analysis**

The AFM technique (MultiView 4000, Nanonics Imaging Ltd., Israel) was employed to measure the thickness of the as-grown and transferred 1L-$MoS_2$. Moreover, the as-collected AFM data is processed by using WSxM 5.0 software for calculating the roughness in the film. The surface morphology of the as-grown, wet-polymeric transferred 1L-$MoS_2$ was carried out using FESEM (Supra 55, Zeiss, Germany). In addition, FESEM images of a few FET devices also carried out.

**RBS characterization**

The compositional analysis of wet-polymeric transferred 1L-$MoS_2$ was carried out using Rutherford backscattering (RBS) spectrometry technique in the carbon resonance (CR) mode. CR-RBS operates on 1.7 MeV tandetron accelerator with 4.27 MeV $4He^{2+}$ beam of 10 micro coulomb charge having energy resolution of 20 keV.

**FET fabrication on two transferred films**

The back-gated FETs were fabricated on wet polymeric transferred 1L-$MoS_2$ film with a channel length of 10 μm. The detailed fabrication protocols of FET can be found elsewhere.[12] The transfer and output FET characteristics were carried out independently.

Multiple FET measurements are carried out for statistical analysis and to ensure the stability of FET measurements.

## Acknowledgements

We thank Dr. Santanu Kumar Parida and Dr. Gopinath Sahoo for their valuable suggestions during the film transfer work. We are thankful to Dr. Sundarvel, IGCAR for the analysis of CR-RBS data.

## References

[1] L. Zhang, Z. Lu, Y. Song, L. Zhao, B. Bhatia, K.R. Bagnall, E.N. Wang, Thermal Expansion Coefficient of Monolayer Molybdenum Disulfide Using Micro-Raman Spectroscopy, Nano letters, 19 (2019) 4745-4751.
[2] A. Sebastian, R. Pendurthi, T.H. Choudhury, J.M. Redwing, S. Das, Benchmarking monolayer MoS 2 and WS 2 field-effect transistors, Nature Communications, 12 (2021) 1-12.
[3] Y.H. Lee, X.Q. Zhang, W. Zhang, M.T. Chang, C.T. Lin, K.D. Chang, Y.C. Yu, J.T.W. Wang, C.S. Chang, L.J. Li, Synthesis of large-area MoS2 atomic layers with chemical vapor deposition, Advanced materials, 24 (2012) 2320-2325.
[4] M. Sharma, A. Singh, R. Singh, Monolayer MoS2 Transferred on Arbitrary Substrates for Potential Use in Flexible Electronics, ACS Applied Nano Materials, 3 (2020) 4445-4453.
[5] S. Fan, Q.A. Vu, M.D. Tran, S. Adhikari, Y.H. Lee, Transfer assembly for two-dimensional van der Waals heterostructures, 2D Materials, 7 (2020) 022005.
[6] D. Akinwande, C. Huyghebaert, C.-H. Wang, M.I. Serna, S. Goossens, L.-J. Li, H.-S.P. Wong, F.H. Koppens, Graphene and two-dimensional materials for silicon technology, Nature, 573 (2019) 507-518.
[7] M.C. Lemme, D. Akinwande, C. Huyghebaert, C. Stampfer, 2D materials for future heterogeneous electronics, Nature communications, 13 (2022) 1392.
[8] A.J. Watson, W. Lu, M.H. Guimarães, M. Stöhr, Transfer of large-scale two-dimensional semiconductors: challenges and developments, 2D Materials, 8 (2021) 032001.
[9] J. Shi, D. Ma, G.-F. Han, Y. Zhang, Q. Ji, T. Gao, J. Sun, X. Song, C. Li, Y. Zhang, Controllable growth and transfer of monolayer MoS2 on Au foils and its potential application in hydrogen evolution reaction, ACS nano, 8 (2014) 10196-10204.
[10] A. Reina, X. Jia, J. Ho, D. Nezich, H. Son, V. Bulovic, M.S. Dresselhaus, J. Kong, Large area, few-layer graphene films on arbitrary substrates by chemical vapor deposition, Nano letters, 9 (2009) 30-35.
[11] Z. Cheng, Q. Zhou, C. Wang, Q. Li, C. Wang, Y. Fang, Toward intrinsic graphene surfaces: a systematic study on thermal annealing and wet-chemical treatment of SiO2-supported graphene devices, Nano letters, 11 (2011) 767-771.
[12] C.A. Bhuyan, K.K. Madapu, K. Prabakar, A. Das, S. Polaki, S.K. Sinha, S. Dhara, A Novel Methodology of Using Nonsolvent in Achieving Ultraclean Transferred Monolayer MoS2, Advanced Materials Interfaces, 9 (2022) 2200030.
[13] A. Pirkle, J. Chan, A. Venugopal, D. Hinojos, C. Magnuson, S. McDonnell, L. Colombo, E. Vogel, R. Ruoff, R. Wallace, The effect of chemical residues on the physical and electrical properties of chemical vapor deposited graphene transferred to SiO2, Applied Physics Letters, 99 (2011) 122108.
[14] J. Liang, K. Xu, B. Toncini, B. Bersch, B. Jariwala, Y.C. Lin, J. Robinson, S.K. Fullerton-Shirey, Impact of Post-Lithography Polymer Residue on the Electrical Characteristics of MoS2 and WSe2 Field Effect Transistors, Advanced Materials Interfaces, 6 (2019) 1801321.
[15] S.E. Panasci, E. Schilirò, G. Greco, M. Cannas, F.M. Gelardi, S. Agnello, F. Roccaforte, F. Giannazzo, Strain, doping, and electronic transport of large area monolayer MoS2 exfoliated on gold and transferred to an insulating substrate, ACS Applied Materials & Interfaces, 13 (2021) 31248-31259.

[16] H.J. Conley, B. Wang, J.I. Ziegler, R.F. Haglund Jr, S.T. Pantelides, K.I. Bolotin, Bandgap engineering of strained monolayer and bilayer MoS2, Nano letters, 13 (2013) 3626-3630.
[17] X. Zhang, Z. Shao, X. Zhang, Y. He, J. Jie, Surface charge transfer doping of low-dimensional nanostructures toward high-performance nanodevices, Advanced Materials, 28 (2016) 10409-10442.
[18] T. Liang, S. Xie, W. Fu, Y. Cai, C. Shanmugavel, H. Iwai, D. Fujita, N. Hanagata, H. Chen, M. Xu, Synthesis and fast transfer of monolayer MoS2 on reusable fused silica, Nanoscale, 9 (2017) 6984-6990.
[19] F. Carrascoso, D.-Y. Lin, R. Frisenda, A. Castellanos-Gomez, Biaxial strain tuning of interlayer excitons in bilayer MoS 2, Journal of Physics: Materials, DOI (2019).
[20] D. Lloyd, X. Liu, J.W. Christopher, L. Cantley, A. Wadehra, B.L. Kim, B.B. Goldberg, A.K. Swan, J.S. Bunch, Band gap engineering with ultralarge biaxial strains in suspended monolayer MoS2, Nano letters, 16 (2016) 5836-5841.
[21] M. Amani, D.-H. Lien, D. Kiriya, J. Xiao, A. Azcatl, J. Noh, S.R. Madhvapathy, R. Addou, K. Santosh, M. Dubey, Near-unity photoluminescence quantum yield in MoS2, Science, 350 (2015) 1065-1068.
[22] X. Liu, K. Huang, M. Zhao, F. Li, H. Liu, A modified wrinkle-free MoS2 film transfer method for large area high mobility field-effect transistor, Nanotechnology, 31 (2019) 055707.
[23] D. Sarkar, X. Xie, J. Kang, H. Zhang, W. Liu, J. Navarrete, M. Moskovits, K. Banerjee, Functionalization of transition metal dichalcogenides with metallic nanoparticles: implications for doping and gas-sensing, Nano letters, 15 (2015) 2852-2862.
[24] T. Kawanago, S. Oda, Utilizing self-assembled-monolayer-based gate dielectrics to fabricate molybdenum disulfide field-effect transistors, Applied Physics Letters, 108 (2016) 041605.
[25] Y. Yang, X. An, M. Kang, F. Guo, L. Zhang, Q. Wang, D. Sun, Y. Liao, Z. Yang, Z. Lei, Distinctive MoS 2-MoP nanosheet structures anchored on N-doped porous carbon support as a catalyst to enhance the electrochemical hydrogen production, New Journal of Chemistry, 45 (2021) 14042-14049.
[26] D. Kiriya, M. Tosun, P. Zhao, J.S. Kang, A. Javey, Air-stable surface charge transfer doping of MoS2 by benzyl viologen, Journal of the American Chemical Society, 136 (2014) 7853-7856.
[27] H. Fang, M. Tosun, G. Seol, T.C. Chang, K. Takei, J. Guo, A. Javey, Degenerate n-doping of few-layer transition metal dichalcogenides by potassium, Nano letters, 13 (2013) 1991-1995.
[28] H. Li, Q. Zhang, C.C.R. Yap, B.K. Tay, T.H.T. Edwin, A. Olivier, D. Baillargeat, From bulk to monolayer MoS2: evolution of Raman scattering, Advanced Functional Materials, 22 (2012) 1385-1390.
[29] Y. Wang, C. Cong, C. Qiu, T. Yu, Raman spectroscopy study of lattice vibration and crystallographic orientation of monolayer MoS2 under uniaxial strain, small, 9 (2013) 2857-2861.
[30] C.A. Bhuyan, K.K. Madapu, S. Dhara, Excitation-dependent photoluminescence intensity of monolayer MoS2: Role of heat-dissipating area and phonon-assisted exciton scattering, Journal of Applied Physics, 132 (2022) 204303.
[31] K.-K. Liu, W. Zhang, Y.-H. Lee, Y.-C. Lin, M.-T. Chang, C.-Y. Su, C.-S. Chang, H. Li, Y. Shi, H. Zhang, Growth of large-area and highly crystalline MoS2 thin layers on insulating substrates, Nano letters, 12 (2012) 1538-1544.
[32] A. Gurarslan, Y. Yu, L. Su, Y. Yu, F. Suarez, S. Yao, Y. Zhu, M. Ozturk, Y. Zhang, L. Cao, Surface-energy-assisted perfect transfer of centimeter-scale monolayer and few-layer MoS2 films onto arbitrary substrates, ACS nano, 8 (2014) 11522-11528.
[33] W.H. Chae, J.D. Cain, E.D. Hanson, A.A. Murthy, V.P. Dravid, Substrate-induced strain and charge doping in CVD-grown monolayer MoS2, Applied Physics Letters, 111 (2017) 143106.
[34] Z. Lin, Y. Zhao, C. Zhou, R. Zhong, X. Wang, Y.H. Tsang, Y. Chai, Controllable growth of large–size crystalline MoS 2 and resist-free transfer assisted with a Cu thin film, Scientific reports, 5 (2015) 1-10.
[35] J.-H. Park, S.H. Choi, W.U. Chae, B. Stephen, H.K. Park, W. Yang, S.M. Kim, J.S. Lee, K.K. Kim, Effective characterization of polymer residues on two-dimensional materials by Raman spectroscopy, Nanotechnology, 26 (2015) 485701.
[36] D.-H. Lien, J.S. Kang, M. Amani, K. Chen, M. Tosun, H.-P. Wang, T. Roy, M.S. Eggleston, M.C. Wu, M. Dubey, Engineering light outcoupling in 2D materials, Nano letters, 15 (2015) 1356-1361.

[37] R. Rao, A.E. Islam, S. Singh, R. Berry, R.K. Kawakami, B. Maruyama, J. Katoch, Spectroscopic evaluation of charge-transfer doping and strain in graphene/MoS 2 heterostructures, Physical Review B, 99 (2019) 195401.

[38] F. Lee, M. Tripathi, R.S. Salas, S.P. Ogilvie, A.A. Graf, I. Jurewicz, A.B. Dalton, Localised strain and doping of 2D materials, Nanoscale, DOI (2023).

[39] E. Mercado, J. Anaya, M. Kuball, Impact of polymer residue level on the in-plane thermal conductivity of suspended large-area graphene sheets, ACS Applied Materials & Interfaces, 13 (2021) 17910-17919.

[40] K.K. Madapu, S. Dhara, Laser-induced anharmonicity vs thermally induced biaxial compressive strain in mono-and bilayer MoS2 grown via CVD, AIP Advances, 10 (2020) 085003.

[41] K.K. Madapu, C.A. Bhuyan, S. Srivastava, S. Dhara, A novel mechanism for understanding the strong enhancement of photoluminescence quantum yield in large-area monolayer MoS 2 grown by CVD, Journal of Materials Chemistry C, 9 (2021) 3578-3588.

[42] Y. Yu, Y. Yu, C. Xu, Y.Q. Cai, L. Su, Y. Zhang, Y.W. Zhang, K. Gundogdu, L. Cao, Engineering substrate interactions for high luminescence efficiency of transition-metal dichalcogenide monolayers, Advanced Functional Materials, 26 (2016) 4733-4739.

[43] K.F. Mak, K. He, C. Lee, G.H. Lee, J. Hone, T.F. Heinz, J. Shan, Tightly bound trions in monolayer MoS 2, Nature materials, 12 (2013) 207-211.

[44] X. Gan, D. Englund, D. Van Thourhout, J. Zhao, 2D materials-enabled optical modulators: From visible to terahertz spectral range, Applied Physics Reviews, 9 (2022).

[45] Y. Lin, X. Ling, L. Yu, S. Huang, A.L. Hsu, Y.-H. Lee, J. Kong, M.S. Dresselhaus, T. Palacios, Dielectric screening of excitons and trions in single-layer MoS2, Nano letters, 14 (2014) 5569-5576.

[46] M. Sojková, Š. Chromik, A. Rosová, E. Dobročka, P. Hutár, D. Machajdík, A. Kobzev, M. Hulman, MoS2 thin films prepared by sulfurization, Nanoengineering: Fabrication, Properties, Optics, and Devices XIV, SPIE, 2017, pp. 218-224.

[47] A. Sebastian, R. Pendurthi, T.H. Choudhury, J.M. Redwing, S. Das, Benchmarking monolayer MoS2 and WS2 field-effect transistors, Nature communications, 12 (2021) 693.

[48] N.B. Shinde, B.D. Ryu, K. Meganathan, B. Francis, C.-H. Hong, S. Chandramohan, S.K. Eswaran, Large-scale atomically thin monolayer 2H-MoS2 field-effect transistors, ACS Applied Nano Materials, 3 (2020) 7371-7376.

[49] E.J. Telford, A. Benyamini, D. Rhodes, D. Wang, Y. Jung, A. Zangiabadi, K. Watanabe, T. Taniguchi, S. Jia, K. Barmak, Via method for lithography free contact and preservation of 2D materials, Nano letters, 18 (2018) 1416-1420.

[50] H. Yang, S. Cai, Y. Zhang, D. Wu, X. Fang, Enhanced electrical properties of lithography-free fabricated MoS2 field effect transistors with chromium contacts, The Journal of Physical Chemistry Letters, 12 (2021) 2705-2711.

[51] K. Haase, F. Talnack, S. Donnhäuser, A. Tahn, M. Löffler, M. Hambsch, S.B. Mannsfeld, A simple lithography-free approach for the fabrication of top-contact OFETs with sub-micrometer channel length, Organic Electronics, DOI (2023) 106819.